\documentclass[letter]{aa}

\usepackage{graphicx}
\usepackage{color}
\usepackage{txfonts}
\begin{document} 

	\title{The eROSITA Final Equatorial-Depth Survey (eFEDS):}
	\subtitle{An X-ray bright, extremely luminous infrared galaxy at $z$ = 1.87}
   	\author{Yoshiki Toba \inst{1,2,3},
   		   	Marcella Brusa \inst{4,5},
			Teng Liu \inst{6},
		   	Johannes Buchner \inst{6},
			Yuichi Terashima \inst{3},	
			Tanya Urrutia \inst{7},
			Mara Salvato \inst{6},	   
			Masayuki Akiyama \inst{8},
			Riccardo Arcodia \inst{6},	
			Andy D. Goulding \inst{9},		
			Yuichi Higuchi \inst{10},	
			Kaiki T. Inoue \inst{10},	
			Toshihiro Kawaguchi \inst{11},
			Georg Lamer \inst{7},		
			Andrea Merloni \inst{6},			
			Tohru Nagao \inst{3},
			Yoshihiro Ueda \inst{1},
			Kirpal Nandra \inst{6}
          }

   \institute{Department of Astronomy, Kyoto University, Kitashirakawa-Oiwake-cho, Sakyo-ku, Kyoto 606-8502, Japan\\
            \email{toba@kusastro.kyoto-u.ac.jp} 
            \and
			Academia Sinica Institute of Astronomy and Astrophysics, 11F of Astronomy-Mathematics Building, AS/NTU, No.1, Section 4, Roosevelt Road, Taipei 10617, Taiwan 
			\and
			Research Center for Space and Cosmic Evolution, Ehime University, 2-5 Bunkyo-cho, Matsuyama, Ehime 790-8577, Japan 		
           	\and
 			 Dipartimento di Fisica e Astronomia, Universit\`a di Bologna, via Gobetti 93/2, 40129 Bologna, Italy
			\and
			INAF- Osservatorio di Astrofisica e Scienza dello Spazio di Bologna, via Gobetti 93/3, 40129 Bologna, Italy
           	\and
           	Max-Planck-Institut f\"ur Extraterrestrische Physik (MPE), Giessenbachstrasse 1, 85748 Garching bei M\"unchen, Germany
           	\and
           	Leibniz-Institut f\"ur Astrophysik, Potsdam (AIP), An der Sternwarte 16, 14482 Potsdam, Germany
           	\and
           	Astronomical Institute, Tohoku University, 6-3 Aramaki, Aoba-ku, Sendai, Miyagi 980-8578, Japan
           	\and
           	Department of Astrophysical Sciences, Princeton University, Princeton, NJ 08540, USA
           	\and
		   	Faculty of Science and Engineering, Kindai University, Higashi-Osaka, 577-8502, Japan
		   	\and
		   	Department of Economics, Management and Information Science, Onomichi City University, Hisayamada 1600-2, Onomichi, Hiroshima 722-8506, Japan
}
\date{\today}

 
\abstract{
In this study, we investigate the X-ray properties of WISE J090924.01+000211.1 (WISEJ0909+0002), an extremely luminous infrared (IR) galaxy (ELIRG) at $z_{\rm spec}$= 1.871 in the {\it eROSITA} final equatorial depth survey (eFEDS).
WISEJ0909+0002 is a WISE 22 $\mu$m source, located in the GAMA-09 field, detected by {\it eROSITA} during the performance and verification phase.
The corresponding optical spectrum indicates that this object is a type-1 active galactic nucleus (AGN).
Observations from {\it eROSITA} combined with {\it Chandra} and {\it XMM-Newton} archival data indicate a very luminous ($L$ (2--10 keV) = ($2.1 \pm 0.2) \times 10^{45}$ erg s$^{-1}$) unobscured AGN with a power-law photon index of $\Gamma$ = 1.73$_{-0.15}^{+0.16}$, and an absorption hydrogen column density of $\log\,(N_{\rm H}/{\rm cm}^{-2}) < 21.0$.
The IR luminosity was estimated to be $L_{\rm IR}$ = (1.79 $\pm$ 0.09) $\times 10^{14}\, L_{\sun}$ from spectral energy distribution modeling based on 22 photometric data (X-ray to far-IR) with {\tt X-CIGALE}, which confirmed that WISEJ0909+0002 is an ELIRG. 
A remarkably high $L_{\rm IR}$ despite very low $N_{\rm H}$ would indicate that we are witnessing a short-lived phase in which hydrogen gas along the line of sight is blown outwards, whereas warm and hot dust heated by AGNs still exist.
As a consequence of {\it eROSITA} all-sky survey, $6.8_{-5.6}^{+16}\times 10^2$ such X-ray bright ELIRGs are expected to be discovered in the entire extragalactic sky ($|b| > 10\degr$).
This can potentially be the key population to constrain the bright-end of IR luminosity functions.
}

   \keywords{Galaxies: active --
             X-rays: galaxies --
             Infrared: galaxies --
             quasars: individual: WISE J090924.01+000211.1
               }

  	\titlerunning{{\it eROSITA} view of an ELIRG in the eFEDS.}
	\authorrunning{Y.Toba et al.}

   \maketitle

%

\section{Introduction}

Galaxies with infrared (IR) luminosity ($L_{\rm IR}$\footnote{Empirically, $L_{\rm IR}$ is defined as the luminosity integrated over a wavelength range of 8--1000 $\mu$m \citep[e.g.,][]{Sanders,Chary}. However, since this definition includes contributions from stellar emissions, we do not adopt any boundary for the integration range on the wavelength. Instead, we employed a physically oriented approach to estimate $L_{\rm IR}$ in this work (see Section \ref{s_LIR}).}) greater than 10$^{13}$$L_{\sun}$ and 10$^{14}$$L_{\sun}$ have been termed as hyper-luminous IR galaxies \citep[HyLIRGs;][]{Rowan-Robinson} and extremely-luminous IR galaxies \citep[ELIRGs:][]{Tsai}, respectively.
The IR luminosity can arise from active galactic nuclei (AGN) and star formation (SF) activity.
During galactic merger events, these luminous IR galaxy populations may correspond to a phase in which the AGN and SF activity reach a peak, shrouded by dense gas and dust clouds \citep[e.g.,][]{Hopkins,Narayanan,Toba_15,Blecha}.
Systematic investigations on Hy/ELIRGs are required to cement the understanding on the co-evolution of such galaxies and their supermassive black holes (SMBHs) during the peak of AGN and SF activity.

Since the advent of IR satellites such as the {\it Wide-field Infrared Survey Explorer} \citep[{\it WISE};][]{Wright} and the {\it Herschel Space Observatory} \citep{Pilbratt}, a large number of Hy/ELIRGs have been discovered \citep[e.g.,][]{Casey,Weedman,Leipski,Toba_16,Duras17}.
However, accurate characterization of the properties of ELIRGs based on multi-wavelength spectral energy distribution (SED) is limited \citep[see e.g.,][]{Toba_18,Toba_20b}.
In particular, X-ray properties of ELIRGs such as hydrogen column density ($N_{\rm H}$) and absorption-corrected X-ray luminosity are poorly understood.

The {\it extended ROentgen Survey with an Imaging Telescope Array} \citep[{\it eROSITA};][]{Merloni,Predehl} has recently probed the X-ray properties of ELIRGs.
{\it eROSITA} is the primary instrument on the Spectrum-Roentgen-Gamma (SRG) mission, which was successfully launched on July 13, 2019.
Since ELIRGs are a spatially rare population and may be X-ray faint due to the obscuration, the {\it eROSITA} all-sky survey with high X-ray sensitivity is an ideal platform to investigate the X-ray properties of ELIRGs.
In this letter, we report the {\it eROSITA} view of ELIRG at $z_{\rm spec}$ = 1.871, WISE J090924.01+000211.1 (hereafter WISEJ0909+0002), in the GAMA-09 field observed by {\it eROSITA}. 
We employed the performance and verification (PV) phase program called {\it eROSITA} Final Equatorial Depth Survey (eFEDS: Brunner et al. in prep.) in our observation.
The eFEDS catalog contains approximately 28,000 X-ray point sources detected over an area of  140 deg$^2$ in a single broad band, with a 5$\sigma$ sensitivity of $f_{\rm 0.3-2.3~keV} \sim 9 \times 10^{-15}$ erg s$^{-1}$ cm$^{-2}$.
We discuss the AGN and its host properties of WISEJ0909+0002 based on SED fitting and X-ray spectral analysis.
Throughout this letter, the adopted cosmology is a flat universe with $H_0$ = 70 km s$^{-1}$ Mpc$^{-1}$, $\Omega_M$ = 0.3, and $\Omega_{\Lambda}$ = 0.7, and the initial mass function of \cite{Chabrier} is assumed.

\section{Data and analysis}
\label{DA}

\subsection{The candidate ELIRG WISEJ0909+0002}
\label{SS}

WISEJ0909+0002 is an ELIRG candidate detected by {\it eROSITA} with a positional uncertainty of 2.0\arcsec.
It is the only ELIRG candidate detected by {\it eROSITA} in a sample of $\sim$ 300 {\it WISE} 22 $\mu$m-selected sources\footnote{The corresponding signal-to-noise ratio of the flux density at 22 $\mu$m is greater than 5.0. Possible stars and artifacts were removed from the sample.} in the eFEDS area with multi-wavelength data ($\sim$50 deg$^2$) (Toba et al. in prep.).
This object is included in the eFEDS X-ray point source catalog with optical-to-mid-IR (MIR) counterparts (Salvato et al. in prep.) called eFEDSJ090924.0+000209.9.
The optical-MIR counterparts were identified using data from the DESI Legacy Imaging Surveys Data Release 8, \citep[LS8:][]{Dey19} which includes $g$, $r$, and $z$ data from the Dark Energy Camera Legacy Survey (DECaLS), and 3.4, 4.6, 12, and 22 $\mu$m data from {\it unWISE} \citep{Lang,Lang16}.
The optical-MIR counterparts are determined using a Bayesian statistics based algorithm ({\tt NWAY} \citealt{Salvato}) and the maximum likelihood method (see Salvato et al. for details), and the multi-wavelength data were compiled (see Toba et al. for details). 
First, we used absorption-corrected fluxes in the 0.5--2.0 and 2.0--10 keV bands obtained via X-ray spectroscopy (see Sect. \ref{Xana}).
The UV data were adopted from the {\it Galaxy Evolution Explorer} \citep[{\it GALEX}:][]{Martin,Bianchi}.
Other optical ($u$, $i$, and $Y$) and near-IR (NIR) data ($J$, $H$, and $K_{\rm s}$) were adopted from the kilo-degree survey (KiDS) DR4 \citep{Kuijken}.
The far-IR (FIR) data (100, 160, 250, 350, and 500 $\mu$m) were adopted from the H-ATLAS DR1 \citep{Valiante}.
The photometry of WISEJ0909+0002 is presented in Table \ref{Photo} in Appendix \ref{A_photo}.
We could confirm the classification of WISEJ0909+0002 as an ELIRG from the SED fitting (see Sects. 2.2 and 3.1), and the redshift was determined spectroscopically using the Sloan Digital Sky Survey \citep[SDSS;][]{York} to be $z_{\rm spec}$ = 1.871.
The observed properties of WISEJ0909+0002 are summarized in Table \ref{Table}.

\begin{table}
\caption{Observed Properties of WISEJ0909+0002.}
\label{Table}
\centering
\begin{tabular}{lr}
\hline\hline
WISE  J090924.01+000211.1		&								\\
\hline
R.A. (SDSS) [hh:mm:ss, J2000.0] 			&	09:09:24.01					\\
Decl. (SDSS) [dd:mm:ss, J2000.0]			&	+00:02:11.06  				\\
Redshift (SDSS)							&	1.871 $\pm$ 0.0001	\\
\hline
SED fitting with {\sf X-CIGALE} (Sect. \ref{s_LIR}) &									\\
\hline
$E(B-V)_{*}$									&	0.13					\\
$M_*$ [$M_{\sun}$]								&	(4.94 $\pm$ 1.39) $\times 10^{11}$	\\
SFR [$M_{\sun}$ yr$^{-1}$]						&	(3.85 $\pm$ 0.06) $\times 10^3$ 	\\
$L_{\rm IR}$ [$L_\sun$]							&	(1.79 $\pm$ 0.09) $\times 10^{14}$	\\
$\nu L_\nu$ (6 $\mu$m) [erg s$^{-1}$]			&	(3.53 $\pm$ 0.80) $\times 10^{46}$	\\
\hline
X-ray spectral analysis	  (Sect. \ref{S_X}) &											\\
\hline
$\log\,(N_{\rm H}/{\rm cm}^{-2})$               &	$< 21.0$							\\
power-law slope ($\Gamma$)                      &	1.73$_{-0.15}^{+0.16}$				\\
$L_{\rm 2-10\ keV}$	[erg s$^{-1}$]				&	$ (2.1 \pm 0.2) \times 10^{45}$ 	\\
\hline
BH properties	  (Sect. \ref{S_X}) &													\\
\hline
$L_{\rm bol}$ [erg s$^{-1}$]						&	($4.3 \pm 0.6)\times10^{47}$	\\
$M_{\rm BH}$ [$M_{\sun}$] 							& 	($7.4 \pm 0.3)\times 10^9$		\\ 
$\lambda_{\rm Edd}$									&	$0.4 \pm 0.1$	\\
$\kappa_{\rm X}$ (=$L_{\rm bol}/L_{\rm 2-10\ keV})$	&	$(2.0 \pm 0.3) \times 10^2$	\\
\hline
\end{tabular}
\end{table}

\subsection{SED fitting with {\tt X-CIGALE}}
\label{s_SED}

To estimate the IR luminosity with high precision, we conducted SED fitting by considering the energy balance between the UV/optical and IR ranges.
We employed the new version of the Code Investigating GALaxy Emission \citep[{\tt CIGALE}; ][]{Burgarella,Noll,Boquien}
called {\tt X-CIGALE}\footnote{\url{https://gitlab.lam.fr/gyang/cigale/tree/xray}} \citep{Yang}, which enables SED fitting from the X-ray to radio range.
In {\tt X-CIGALE}, we are able to handle many parameters, such as the star formation history (SFH), single stellar population (SSP), attenuation law, AGN emission, dust emission, radio synchrotron emission, and AGN X-ray emission \cite[see e.g.,][]{Boquien14,Boquien16,Buat,LoFaro,Toba_19b,Toba_20c}. 
A brief description of each parameter is provided in Appendix \ref{Ap_SED} and the parameter ranges used in the SED fitting are presented in Table \ref{Param} \citep[see also][and references therein]{Boquien,Yang}.

     \begin{figure*}
   \centering
   \includegraphics[width=0.7\textwidth]{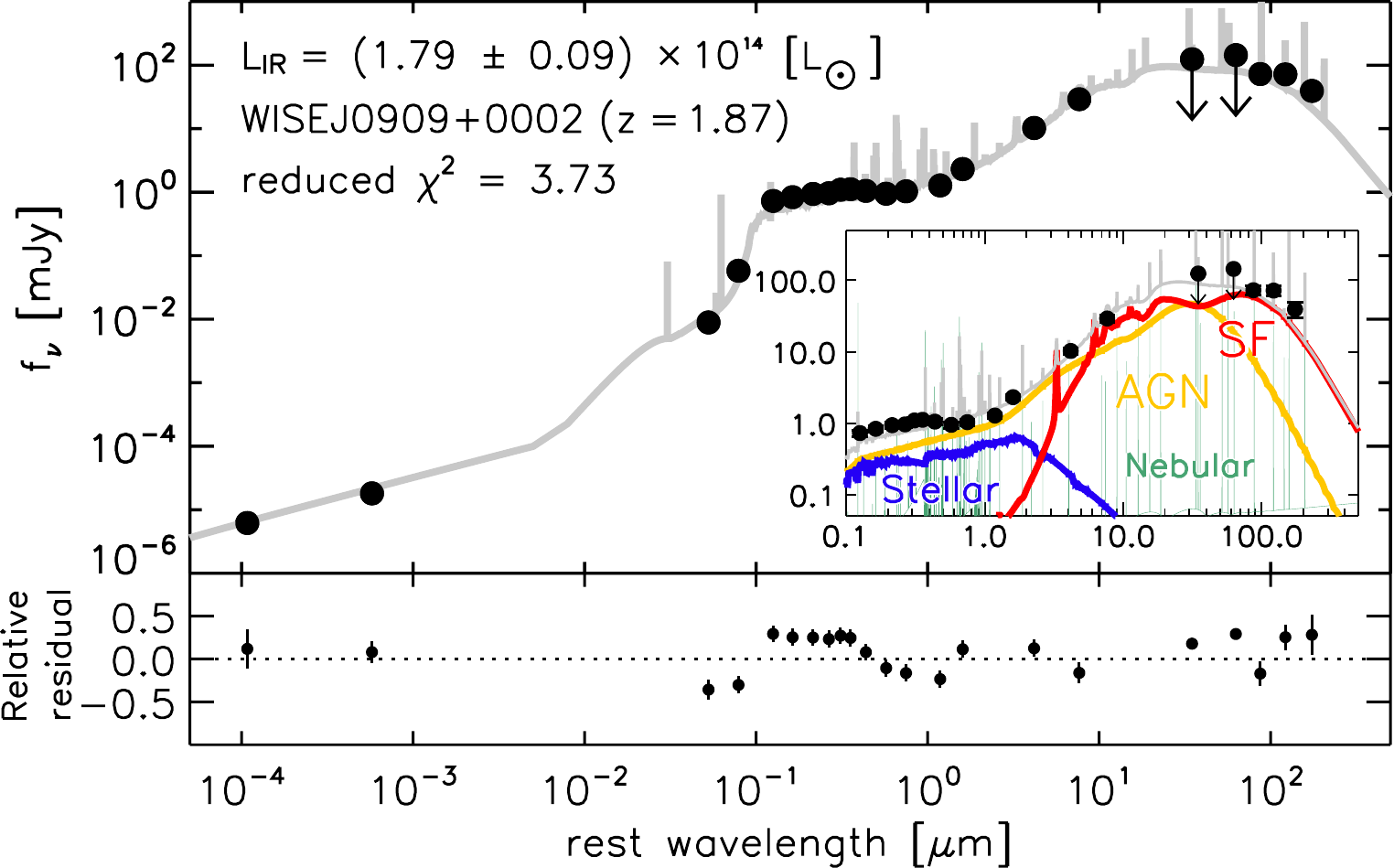}
   \caption{The best-fit SED of WISEJ0909+0002. The black points are photometric data, and the gray solid line represents the resultant best-fit SED. The inset figure shows the SED at 0.1--500 $\mu$m, where the contributions from the stellar, nebular, AGN, and SF components to the total SED are shown as blue, green, yellow, and red lines, respectively. The relative residual (defined as the best-fit value minus data with respect to data) are shown at the bottom, where the black line represents the case when the residual is zero.}
   \label{SED}
   \end{figure*}

\subsection{X-ray spectral analysis}
\label{Xana}

WISEJ0909+0002 was observed by {\it eROSITA} with a net exposure time of 2.3 ks in 2019.
The {\it eROSITA} X-ray spectrum of WISEJ0909+0002 is extracted using eSASS \texttt{srctool} v1.60\footnote{This software derives source level products from calibrated {\it eROSITA} event files and their ancillary meta-data. See \url{https://erosita.mpe.mpg.de/eROdoc/tasks/srctool_doc.html} for details.} within a circle of 50\arcsec in diameter, and the background spectrum is extracted inside an annulus between 114\arcsec and 623\arcsec, excluding all nearby sources (see Liu et al. in prep., in details).
This source was observed by {\it Chandra} in 2004 (Observation ID = 5703), and {\it XMM-Newton} in 2013 (Observation ID = 0725310143).
We extracted the {\it Chandra} and {\it XMM} spectra to perform a joint fitting of the {\it eROSITA} spectrum in the 0.2--8 keV band (90 net counts), {\it Chandra} spectrum in the 0.5--7 keV (17 net counts), and the {\it XMM} EPIC-pn, EPIC-MOS1, and EPIC-MOS2 spectra in the 0.5--8 keV band (with net counts of 33, 11, and 14, respectively).

The {\it Chandra} data are processed with CIAO 4.12 and CALDB 4.9.0. 
The background is stable during observation with a net exposure time of 1.3 ks.
The source spectrum is extracted from a circular region centered at the WISEJ0909+0002 with a radius of 4$\farcs$9. 
The background spectrum is obtained from an annular region around the source.
The {\it XMM-Newton} data are processed with the science analysis system version 18.0.0, combined with the current calibration files as of June 2019. 
The source spectra are extracted from a circular region with a radius of 17$\farcs$5. 
The background spectra are adopted from a nearby source free region. 
The background was stable during the observation, and the net exposure times for the PN, MOS1, and MOS2 detectors were 2.5, 2.7, and 2.7 ks, respectively.

We model the source with a power-law model with galactic absorption ($N_{\rm H}=2.58\times 10^{20}$ cm$^{-2}$) and intrinsic absorption ({\tt constant*TBabs*zTBabs*powerlaw} in {\tt xspec}). 
Three constant factors (0.65 for {\it Chandra} and 1.36 for {\it XMM}) were used to account for the potential variability and calibration difference between {\it eROSITA}, {\it Chandra}, and {\it XMM} (but see Appendix \ref{Val}).
We also fitted the spectra with the same model, setting all the constant factors to unity, in order to measure the averaged luminosity. Fixing these factors has no impact on the spectral shape parameters, because the {\it XMM} and {\it Chandra} spectra have much lower counts than that of {\it eROSITA}.
To derive constraints on the parameters, we employed the Bayesian method BXA \citep[][]{Buchner2014,Buchner2019}, adopting wide uniform priors for the power-law slope and the constant factors, and log-uniform priors for the column density and normalization. 
In Table \ref{Table}, we list the posterior median and 68\% confidence interval of the parameters, as well as the absorption-corrected rest-frame 2--10 keV luminosity (see also Sect.\ref{S_X}).

\section{Results and discussion}

\subsection{IR luminosity and host properties}
\label{s_LIR}

Fig. \ref{SED} shows the best-fit SED of WISEJ0909+0002 (see also Fig. \ref{image} for multi-wavelength images).
We find that the observed data points (except for the observed-frame optical data) are well explained by the combination of stellar, nebular, AGN, and SF components.
The best-fit SED, particularly of the AGN accretion disk, underestimates the flux densities in the rest-frame 0.1--1 $\mu$m by 0.2--0.3 dex, which induces a relatively large reduced $\chi^2$ (= 3.7).
This is partially due to the fact that {\tt X-CIGALE} fixes a slope of the power-law disk component (i.e., the optical spectral index, $\alpha_{\rm opt}$) to be 0.5 in $\lambda$--$f_{\nu}$ space where $\alpha_{\rm opt}$ is defined in the wavelength rage 0.125--1 $\mu$m (see Equation 9 in \citealt{Yang}).
However, $\alpha_{\rm opt}$ for WISEJ0909+0002 inferred from the SDSS spectrum and NIR data is $\sim 0.2$\footnote{This value is within a dispersion of $\alpha_{\rm opt}$ distribution for SDSS quasars \citep[e.g.,][]{Vanden}.}, which is flatter than that employed in {\tt X-CIGALE}.
The relative residual defined as (data -- best-fit)/data is also plotted in the bottom panel of Fig.\ref{SED}.
In the optical--NIR region, the relative residual is slightly larger at shorter wavelengths, which can support the above interpretation.

The resultant IR luminosity is $L_{\rm IR}$ = (1.79 $\pm$ 0.09) $\times 10^{14}$ $L_{\sun}$, which establishes the classification of WISEJ0909+0002 as an ELIRG.
The AGN fraction defined as $L_{\rm IR}$(AGN)/$L_{\rm IR}$ is $\sim 0.4$, suggesting that $L_{\rm IR}$ for WISEJ0909+0002 may have contributions from AGN and SF.
We note that recent SED fitting codes such as {\tt CIGALE} and Multiwavelength
Analysis of Galaxy Physical Properties \citep[{\tt MAGPHYS};][]{da_Cunha} employ physically-motivated $L_{\rm IR}$ without any boundary for the integration range for the wavelength. 
Hence, {\tt X-CIGALE} purely considers the energy re-emitted by dust that absorbs UV\/optical photons from AGN\/SF to estimate $L_{\rm IR}$ \citep[see also][]{Toba_20b}.

The derived color excess of the stellar emission ($E(B-V)_*$), stellar mass ($M_*$), and SFR output from {\tt X-CIGALE} are $E(B-V)_*= 0.13$, $M_* =$ (4.94 $\pm$ 1.39) $\times 10^{11}\,M_{\sun}$, and SFR = (3.85 $\pm$ 0.06) $\times 10^3$ $M_{\sun}$ yr$^{-1}$, respectively, where the SFR was estimated based only on the resultant parameters of the SFH output by {\tt X-CIGALE} \citep[see][for more details]{Boquien}.
The $M_*$--SFR relation of WISEJ0909+0002 shows a significant positive offset ($\sim$1 dex) with respect to the main-sequence galaxies at 1.5 $< z < 2$ for the same stellar mass \citep{Tomczak,Pearson}.
This indicates that WISEJ0909+0002 can be considered a starburst galaxy.
We note that it is often difficult to decompose the UV/optical SED of quasars into stellar and AGN emission with limited photometric data \citep[see e.g.,][]{Merloni10,Bongiorno,Toba_18}.
Hence, the derived $M_*$ may have a large uncertainty due to possible contamination from emission from the AGN accretion disk.
Recent works have also reported that AGNs with higher luminosities exhibit higher AGN contribution to FIR emission than less luminous AGNs \cite[e.g.][]{Symeonidis16,Symeonidis}.
It has been shown that contribution from the AGN torus to FIR emission is small, and FIR emission can be predominantly linked to dust heated by the AGN at kpc-scales. 
Since the SED modeling in our study does not take into account such dust, the derived $f_{\rm AGN}$ and SFR may be underestimated and overestimated, respectively.
Nevertheless, given the fact that possible underestimated/overestimated values for $M_*$ and SFR for type 1 AGNs are expected to be 0.3 dex  (e.g., \citealt{Symeonidis16}; Toba et al. in prep.), our conclusion (i.e.,  WISEJ0909+0002 can be considered a starburst galaxy) is reasonable.

\subsection{AGN properties}
\label{S_X}

\begin{figure}
   \centering
   \includegraphics[width=0.4\textwidth]{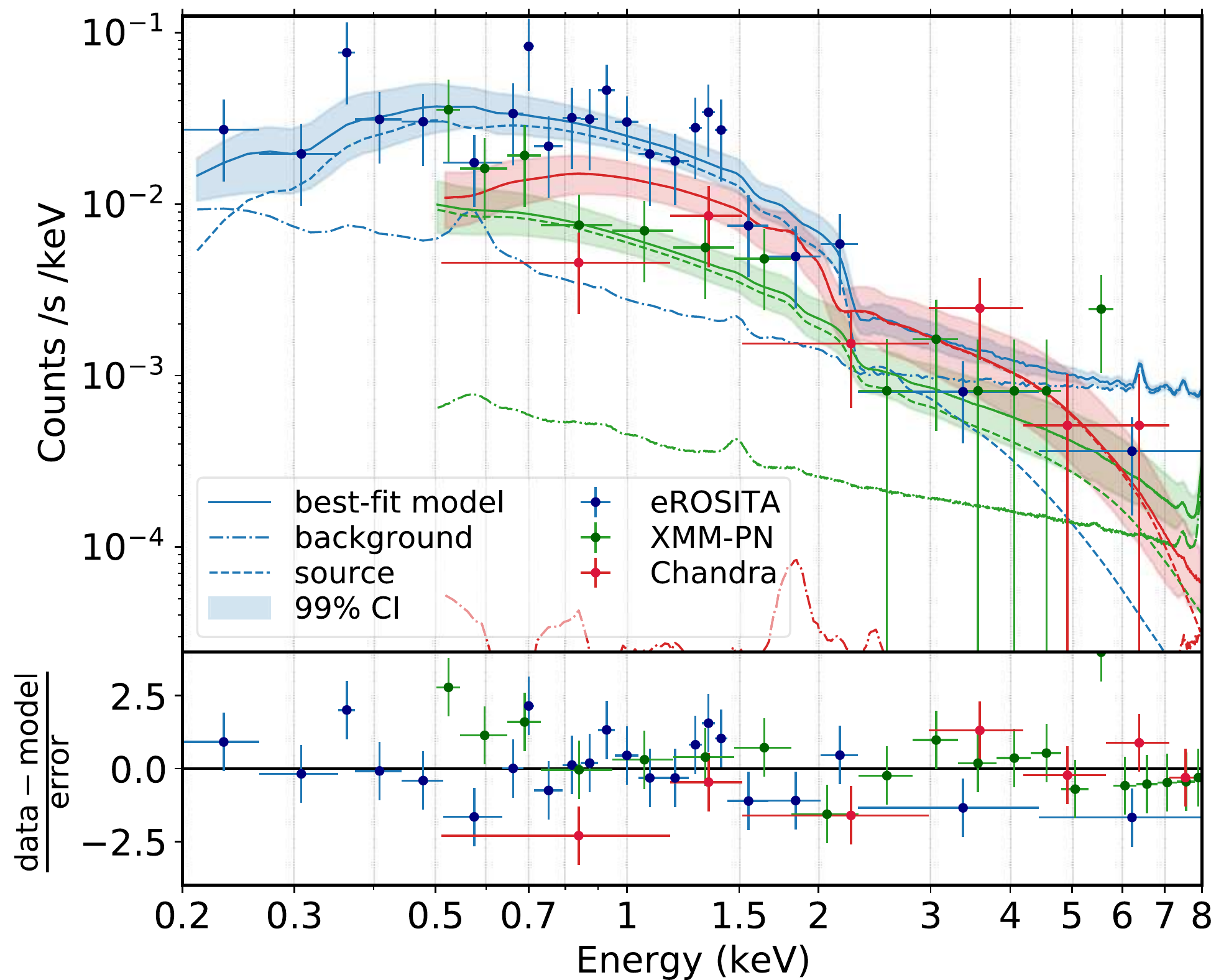}
   \caption{X-ray spectra of {\it eROSITA} (blue), EPIC-pn (green), and {\it Chandra} (red) folded with the energy responses and fitted with the same model (absorbed power-law). The data are binned to a minimum significance of 2$\sigma$ for convenience. The best-fit model and the corresponding 99\% confidence interval are displayed in blue, green, and red for {\it eROSITA}, EPIC-pn, and {\it Chandra}, respectively. The lower panel displays the (data-model)/$\sqrt{{\rm model}}$.}
   \label{Xspec}
   \end{figure}

To the best of our knowledge, this is the first work on {\it eROSITA} data to determine the X-ray properties of an ELIRG directly from the X-ray spectra\footnote{\cite{Krawczyk} reported $L_{\rm 2-10\ keV}$ for WISEJ0909+0002 to be $\log\,(L_{\rm 2-10\ keV}/{\rm erg~s}^{-1}) = 45.45 \pm 0.18$, which is in good agreement with our work. However, their derivation is based on an empirical relation between 2500 \AA \ and the 2 keV luminosity for a given photon index ($\Gamma = 2$).}.
Fig. \ref{Xspec} shows the X-ray spectra of WISEJ0909+0002, in which the data are well fitted using the model described in Sect. \ref{Xana}.
The resulting values of the photon index ($\Gamma$), $N_{\rm H}$, and absorption-corrected hard X-ray luminosity in the rest-frame 2--10 keV ($L_{\rm 2-10\ keV}$) are $\Gamma$ = 1.73$_{-0.15}^{+0.16}$, $\log\,(N_{\rm H}/{\rm cm}^{-2}) < 21.0$, and $L_{\rm 2-10\ keV} = (2.1 \pm 0.2) \times 10^{45}$ erg s$^{-1}$.
These values are typical for luminous unobscured type-1 AGNs.

WISEJ0909+0002 exhibits broad emission lines of C{\,\sc iv} and Mg{\,\sc ii} with full width at half maximum (FWHM) values of 5580$\pm$102 and 6173$\pm$115 km s$^{-1}$, respectively \citep{Rakshit} in the SDSS spectrum.
\cite{Rakshit} measured the monochromatic luminosity at 3000 \AA \footnote{\cite{Rakshit} decomposed the host galaxy from spectra only for quasars at $z < 0.8$, and thus the contribution from the host galaxy for WISEJ0909+0002 was not taken into account to estimate $L_{\rm 3000}$. But since the host contribution to $L_{\rm 3000}$ for luminous quasars is expected to be negligible, we did not apply for host subtraction \citep[see also][]{Calderone}.}, ($L_{\rm 3000}$ = $7.2 \times 10^{46}$ erg s$^{-1}$) and FWHM of Mg{\,\sc ii}, which contribute to the bolometric luminosity ($L_{\rm bol}$) and BH mass ($M_{\rm BH}$), by using multi-component spectral fitting.
$L_{\rm bol}$ is then calculated using $L_{\rm 3000} \times {\rm BC_{\rm 3000}}$, where BC$_{\rm 3000}$ is bolometric correction \citep[BC$_{\rm 3000}$ = 5.9 $\pm$ 0.8;][]{Nemmen,Runnoe}. $M_{\rm BH}$ is estimated using a single epoch method reported by \cite{Vestergaard}.
The uncertainty in $L_{\rm bol}$ and $M_{\rm BH}$ is calculated through error propagation in the same manner as in \cite{Toba_20d}.

The resultant $L_{\rm bol}$ and $M_{\rm BH}$ are $L_{\rm bol} = (4.3 \pm 0.6)\times 10^{47}$ erg s$^{-1}$ and $M_{\rm BH} = (7.4 \pm 0.3)\times 10^9$ $M_{\sun}$, respectively.
The Eddington ratio ($\lambda_{\rm Edd} = L_{\rm bol}/L_{\rm Edd}$) is determined to be $\lambda_{\rm Edd} = 0.4 \pm 0.1$.
Following \cite{Toba_17}, we estimated $L_{\rm bol}$ by integrating the best-fit SED template of the AGN
component output by {\tt X-CIGALE} over wavelengths longward of Ly$\alpha$.
We obtained $\log\,(L_{\rm bol}^{\rm SED}/{\rm erg \,s^{-1}})\sim47.5$, which is consistent with $L_{\rm bol}$ derived from the SDSS spectrum within the error.

It has been reported that there is a strong luminosity dependence of the hard X-ray bolometric correction $\kappa_{\rm X}$ (=$L_{\rm bol}/L_{\rm 2-10\ keV})$ at high luminosities \citep[e.g.,][]{Lusso10,Duras}.
The resulting $\kappa_{\rm X}$ of WISEJ0909+0002 is $(2.0 \pm 0.3) \times 10^2$, which is consistent with the  value obtained via an empirical relationship between $\kappa_{\rm X}$ and $L_{\rm bol}$ for type-1 AGNs ($\kappa_{\rm X} \sim 207$\footnote{If we employ relationships of $\kappa_{\rm X}$--$\lambda_{\rm Edd}$ and $\kappa_{\rm X}$--$M_{\rm BH}$ presented in \cite{Duras}, the resultant $\kappa_{\rm X}$ values are $\sim$36 and $\sim$116, respectively. Although $\lambda_{\rm Edd}$-based $k_{\rm X}$ is significantly smaller than what is reported ($\sim$200), it would be hard to discuss here the origin of this discrepancy, given a huge scatter (0.3--0.4 dex) in those empirical relationships. We note that the $\kappa_{\rm X}$--$\lambda_{\rm Edd}$ and $\kappa_{\rm X}$--$M_{\rm BH}$ relations in \cite{Duras} are based not only on type 1 but also on type 2 AGNs although $k_{\rm X}$--$L_{\rm bol}$ relation is derived based only from type 1 AGNs, which might induce the discrepancy.}) with a dispersion of 0.26 dex, reported by \cite{Duras}.
$\kappa_{\rm X}$ is also correlated with $M_{\rm BH}$ and $\lambda_{\rm Edd}$ \citep[e.g.,][]{Vasudevan,Lusso12,Martocchia}.
The measured value of $\kappa_{\rm X}$ is larger than the values estimated from empirical relations for type-1 AGNs of (i) $\kappa_{\rm X}$--$M_{\rm BH}$ ($\sim$113) and (ii) $\kappa_{\rm X}$--$\lambda_{\rm Edd}$ ($\sim$153) reported in \cite{Martocchia} and \cite{Lusso12}, respectively, but still consistent with them within the scatters of the correlations.
We conclude that the observed AGN properties ($M_{\rm BH}$, $L_{\rm bol}$, $L_{\rm X}$, and $\lambda_{\rm Edd}$) of WISEJ0909+0002 follow empirical relations for type-1 AGNs reported previously.

\subsection{Expected surface number density of X-ray bright ELIRGs}
\label{suface}

We then discuss the number of X-ray bright ELIRGs expected to be discovered as a consequence of the {\it eROSITA} all-sky survey (eRASS) that will continue until the end of 2023 \citep{Predehl}.
Following \cite{Toba_16}, we estimate the surface number density by taking into account the survey area addressed in this work and the detection completeness of the {\it WISE} and {\it eROSITA} all-sky surveys.
The footprint of ELIRG survey in the eFEDS (GAMA-09 field) is currently determined by the overlapping region of KiDS DR4 and H-ATLAS DR1, which is approximately 50 deg$^2$ \citep[e.g.,][]{Fleuren}.
WISEJ0909+0002 is drawn from the 22 $\mu$m-selected sample, and its flux density at 22 $\mu$m ($\sim$30 mJy) is well above the fluxes at the 95\% completeness limit\footnote{\url{https://wise2.ipac.caltech.edu/docs/release/allsky/expsup/sec6_5.html}.} ($\sim$7 mJy).
The X-ray flux in the 0.5--2 keV band of WISEJ0909+0002 is $\sim1.6 \times 10^{-13}$ erg s$^{-1}$ cm$^{-2}$, which is roughly one order of magnitude brighter than that at 95\% completeness ($\sim2 \times 10^{-14}$ erg s$^{-1}$ cm$^{-2}$) \citep{Merloni}.
Therefore, the expected surface number density of X-ray bright ELIRGs is $2.0_{-1.7}^{+4.6} \times 10^{-2}$ deg$^{-2}$, where the 1$\sigma$ confidence limits are estimated based on the Poisson statistics reported by \cite{Gehrels}.
Therefore, $2.0_{-1.7}^{+4.6} \times 10^{-2}$ $\times$ 34,100 deg$^2$ (corresponding to the sky area at $|b| > 10\degr$) $=6.8_{-5.6}^{+16}\times 10^2$ objects are expected to be discovered in the entire extragalactic sky, which suggests that X-ray bright ELIRGs are a rare population, and this phase appears in a significantly short time scale.

\subsection{The evolutionary stage of WISEJ0909+0002}
\label{evo_stage}

A remarkable property of WISEJ0909+0002 is the low $N_{\rm H}$ ($< 10^{21.0}$ cm$^{-2}$), despite the fact that its $L_{\rm IR}$ value is extraordinary large (i.e., this ELIRG is presumably embedded within a large amount of gas and dust).
Furthermore, WISEJ0909+0002 is known as a broad absorption line (BAL) quasar \citep{Trump,Ganguly,Moravec}.
To check the consistency of $N_{\rm H}$ derived from the X-ray spectral analysis, we estimate $N_{\rm H}$ from an expected ionization parameter ($U$) inferred from the C{\,\sc iv}~$\lambda1549$ and N{\,\sc v}~$\lambda1240$ lines as follows.
\cite{Moravec} reported that the lower limits of the column density ($N_{\rm ion}$) of C{\,\sc iv} and the corresponding optical depth of the line center ($\tau_0$) are $2.89\times 10^{15}$ cm$^{-2}$ and 0.55, respectively, while those of $N_{\rm ion}$ and $\tau_0$ for N{\,\sc v} are $5.61\times 10^{15}$ cm$^{-2}$ and 0.7, respectively.
This suggests that (i) relative abundance of N{\,\sc v} (N$^{4+}$) is larger than C{\,\sc iv} (C$^{3+}$) and (ii) those ions are optically-thin.
Under the above conditions\footnote{We are not able to rule out the possibility that the X-ray absorber and BAL gas are not exactly the same, although these absorbers are often reported to have similar properties (e.g., velocity and column density) and are likely to be the same for some quasars \citep{Hamann_18}. A detailed analysis of the X-ray absorption line is required to address this issue, which may be the scope of future work.}, we estimate the lower-limits to $U$ and $N_{\rm H}$ based on photoionization calculations reported by \cite{Hamann} and \cite{Wang}.
As a result, we obtain $\log\,U > -1.2$ and $\log\, N_{\rm H} > 21$ cm$^{-2}$, which is consistent with what is derived by \cite{Xu} for other quasars with a similar BAL feature as WISEJ0909+0002.

The derived $N_{\rm H}$ is larger than the value obtained from the X-ray spectral analysis.
This discrepancy may be attributed to the variability of $N_{\rm H}$.
The BAL feature of WISEJ0909+0002 was reported based on the averaged UV/optical spectra measured in 2001--2010 (the averaged epoch is 2006).
The {\it Chandra}  X-ray spectrum was measured in 2004.
If we derive $N_{\rm H}$ only from {\it Chandra}, we obtain $\log\,(N_{\rm H}/{\rm cm}^{-2}) = 21.5_{-1.63}^{+1.24}$ (see Appendix \ref{Val}), which is roughly consistent with that estimated from the BAL feature.
Since {\it XMM} and {\it eROSITA} observed this object in 2013 and 2019, respectively, it may be possible that $N_{\rm H}$ of WISEJ0909+0002 decreased in 10-15 years.
Given the large uncertainty of {\it Chandra}-based $N_{\rm H}$, it may be hard to conclude that this variability is significant.
Nevertheless, if this drastic decline of $N_{\rm H}$ is true, the BAL feature of the WISEJ0909+0002 spectra is expected to be weak, or even missing in the current UV/optical spectra. 
This is an issue to be addressed in a future work.

Recent galactic merger simulations show that an evolutionary stage with high $L_{\rm IR}$ and low $N_{\rm H}$ appears for $\sim$ 5--10 Myr during the major merger event in which hydrogen gas along the line of sight is blown outwards, whereas warm and hot dust heated by AGNs still exist (Yutani et al. in prep.\footnote{Yutani et al. conducted high-resolution N-body/SPH simulations with {\tt ASURA} \citep{Saitoh_08,Saitoh_09}, and investigated the time evolution of SED of mergers using the the radiative transfer simulation code {\tt RADMC-3D} \citep{Dullemond}.}).
We checked images from the {\it Hubble Space Telescope} ({\it HST}) to see if there are any clues of potential merger event, but no such observation is apparent in the image, as shown in Figure \ref{image}. 
The FWHM and radial profile of WISEJ0909+0002 are comparable to those of possible stars around our object in the same FoV of HST, suggesting that WISEJ0909+0002 can be considered as a point source.
This may be because the merger sign is too weak to be detected currently, given the fact that an object with low $N_{\rm H}$ and high $L_{\rm IR}$ would appear in the late stage of a galactic merger (Yutani et al. in prep.).
In addition, strong nuclear emissions often erase the trace of the merger sign.
However, the optical spectrum exhibits blue wing for some emission lines such as C{\,\sc iii}] \citep{Rakshit}, which may support the idea that this ELIRG is in the blow-out phase.
Thus, we are witnessing a short-lived ELIRG phase in the course of galactic evolution for WISEJ0909+0002.
Owing to the rarity and extreme luminosity, such X-ray bright ELIRGs will be ideal environments to investigate the bright-end of X-ray and IR luminosity functions.

\begin{acknowledgements}

We gratefully acknowledge the anonymous referee for a careful reading of the manuscript and very helpful comments.
We thank Takashi Horiuch, Keiichi Wada, and Naomichi Yutani for useful discussion and comments.

This work is based on data from {\it eROSITA}, the primary instrument aboard SRG, a joint Russian-German science mission supported by the Russian Space Agency (Roskosmos), in the interests of the Russian Academy of Sciences represented by its Space Research Institute (IKI), and the Deutsches Zentrum f\"ur Luft- und Raumfahrt (DLR). 
The SRG spacecraft was built by Lavochkin Association (NPOL) and its subcontractors, and is operated by NPOL with support from the Max-Planck Institute for Extraterrestrial Physics (MPE).
The development and construction of the {\it eROSITA} X-ray instrument was led by MPE, with contributions from the Dr. Karl Remeis Observatory Bamberg \& ECAP (FAU Erlangen-Nuernberg), the University of Hamburg Observatory, the Leibniz Institute for Astrophysics Potsdam (AIP), and the Institute for Astronomy and Astrophysics of the University of T\"ubingen, with the support of DLR and the Max Planck Society. 
The Argelander Institute for Astronomy of the University of Bonn and the Ludwig Maximilians Universit\"at Munich also participated in the science preparation for {\it eROSITA}. 
The {\it eROSITA} data shown here were processed using the eSASS/NRTA software system developed by the German {\it eROSITA} consortium. \\

This work is supported by JSPS KAKENHI grant Nos. 18J01050 and 19K14759 (Y.Toba), 16K05296 (Y.Terashima), 20H01949 (T.Nagao), and 20H01946 (Y.Ueda).

\end{acknowledgements}


\begin{appendix}
\section{Photometry of WISEJ0909+0002}
\label{A_photo}

The multi-band photometry of WISEJ0909+0002 is presented in Table \ref{Photo}.
The X-ray flux densities are corrected for galactic and intrinsic absorption, where $\log, (N_{\rm H}/{\rm cm}^{-2} )= 20.1$ is used for intrinsic absorption correction (see Fig.~\ref{chandra}a).
Flux densities in the optical to MIR are corrected for galactic extinction according to \cite{Fitzpatrick} and \cite{Schlafly}.
The SPIRE FIR flux densities are corrected for flux-boosting \citep[see e.g.,][]{Valiante,Toba_19b}.

\begin{table}[h]
\caption{Photometry of WISEJ0909+0002.}
\label{Photo}
\centering
\begin{tabular}{lc}
\hline \hline
Band	&	Flux density [mJy]					\\
\hline
{\it eROSITA} $f_{\rm 2-10\,keV}$ 		&	$(6.18 \pm 1.30)\times 10^{-6}$	\\
{\it eROSITA} $f_{\rm 0.5-2\,keV}$ 		&	$(18.1 \pm 1.47)\times 10^{-6}$	\\
{\it GALEX}	FUV 						&	$(8.89 \pm 0.54)\times 10^{-3}$	\\
{\it GALEX}	NUV 						&	$(57.8 \pm 1.34)\times 10^{-3}$	\\
KIDS 	$u$-band 						&	$(72.6 \pm 0.03)\times 10^{-2}$	\\
DECaLS 	$g$-band 						&	$(83.2 \pm 0.05)\times 10^{-2}$	\\
DECaLS  $r$-band 						&	$(93.6 \pm 0.05)\times 10^{-2}$	\\
KiDS 	$i$-band 						&	$(96.0 \pm 0.07)\times 10^{-2}$	\\
DECaLS	$z$-band 						&	$(10.9 \pm 0.01)\times 10^{-1}$	\\
KiDS/VIKING $Y$-band 					&	$(11.1 \pm 0.01)\times 10^{-1}$ \\
KiDS/VIKING $J$-band 					&	$(10.5 \pm 0.01)\times 10^{-1}$ \\
KiDS/VIKING $H$-band 					&	$(9.46 \pm 0.02)\times 10^{-1}$ \\
KiDS/VIKING $K_{\rm S}$-band 			&	$(10.3 \pm 0.02)\times 10^{-1}$ \\
{\it unWISE} 3.4 $\mu$m 				&	$(12.8 \pm 0.02)\times 10^{-1}$ \\ 
{\it unWISE} 4.6 $\mu$m 				&	$(23.1 \pm 0.05)\times 10^{-1}$  \\
{\it unWISE} 12  $\mu$m 				&	10.2 $\pm$ 0.13 \\
{\it unWISE} 22  $\mu$m 				&	29.1 $\pm$ 1.88 \\
PACS 100 $\mu$m 						&	$<124$\tablefootmark{a}		\\
PACS 160 $\mu$m 						&	$<144$\tablefootmark{a}		\\
SPIRE 250 $\mu$m 						&	72.5 $\pm$ 7.20	\\
SPIRE 350 $\mu$m 						&	71.6 $\pm$ 8.09	\\
SPIRE 500 $\mu$m 						&	39.2 $\pm$ 8.49	\\
\hline
\end{tabular}
\tablefoot{
\tablefoottext{a}{3$\sigma$ upper limit.}
}
\end{table}

\section{Parameter ranges used in the SED fitting with {\tt X-CIGALE}}
\label{Ap_SED}

Table \ref{Param} summarizes the parameter ranges used in the SED fitting with {\tt X-CIGALE} (see Sect. \ref{s_SED}).
Best-fit values (termed as {\tt BEST\_$parameter name$} in the output file from {\tt X-CIGALE}) are denoted by boldface.
Here, we briefly explain the each parameter for the SED fitting.

We assume the delayed SFH with recent starburst \citep{Ciesla} with parameterizing  e-folding time of the main stellar population model ($\tau_{\rm main}$), age of the main stellar population in the galaxy, age of burst, and ratio of the SFR after and before the burst (R\_sfr).
A starburst attenuation curve \citep{Calzetti,Leitherer} is used for the dust attenuation in which we parameterize color excess of the nebular emission lines ($E(B-V)_{\rm lines}$). 
We chose the SSP model \citep{Bruzual}, assuming the IMF of \cite{Chabrier}, and the standard nebular emission model included in {\tt X-CIGALE} \citep[see ][]{Inoue}.
AGN emission is modeled by using the {\tt SKIRTOR} \citep{Stalevski}.
This torus model consists of 7 parameters; torus optical depth at 9.7 $\mu$m ($\tau_{\rm 9.7}$), torus density radial parameter ($p$), torus density angular parameter ($q$), angle between the equatorial plane and edge of the torus ($\Delta$), ratio of the maximum to minimum radii of the torus ($R_{\rm max}/R_{\rm min}$), the viewing angle ($\theta$), and the AGN fraction in total IR luminosity ($f_{\rm AGN}$). 
Dust grain emission is modeled by \cite{Draine} in which we parameterize the mass fraction of PAHs ($q_{\rm PAH}$), the minimum radiation field ($U_{\rm min}$), the power-low slope of the radiation field distribution ($\alpha$), and the fraction illuminated with a variable radiation field ranging from $U_{\rm min}$ to $U_{\rm max}$ ($\gamma$).
X-ray emission is modeled with fixed power-low photon indices of AGN, low-mass X-ray binaries (LMXB), and high-mass X-ray, binaries (HMXB).

\begin{table}[h]
\caption{Parameter Ranges Used in the SED Fitting with {\tt X-CIGALE}} 
\label{Param}
\centering
\begin{tabular}{lc}
\hline \hline
Parameter & Value\\
\hline
\multicolumn{2}{c}{Delayed SFH with recent starburst \citep{Ciesla}}\\
\hline
$\tau_{\rm main}$ [Gyr] & 1.0, 4.0, 8.0, {\bf 12} \\
age [Gyr] 				& {\bf 0.5}, 1.0, 1.5, 2.0 \\
age of burst [Myr] 		& 10, {\bf 50}, 100 \\
R\_sfr					& 1, 5, {\bf 10} \\
\hline
\multicolumn{2}{c}{SSP \citep{Bruzual}}\\
\hline
IMF				&	\cite{Chabrier} \\
Metallicity		&	0.02  \\
\hline
\multicolumn2c{Nebular emission \citep{Inoue}}\\
\hline
$\log\, U$					&	{\bf $-$3.0}, $-$2.0, $-$1.0	\\
line width [km s$^{-1}$]    &    300	\\
\hline
\multicolumn2c{Dust attenuation \citep{Calzetti,Leitherer}}\\
\hline
$E(B-V)_{\rm lines}$ &  {\bf 0.3}, 0.5, 1.0 \\
\hline
\multicolumn{2}{c}{AGN Emission \citep{Stalevski12,Stalevski}}\\
\hline
$\tau_{\rm 9.7}$ 			&  	3, 7, {\bf 11}	\\
$p$							&	{\bf 0.5},  1.5	\\
$q$							&	{\bf 0.5}, 1.5	\\
$\Delta$ [$\degr$]			&	80				\\
$R_{\rm max}/R_{\rm min}$ 	& 	30 				\\
$\theta$ [$\degr$]			&	{\bf 0}, 10, 20		\\
$f_{\rm AGN}$ 				& 	{\bf 0.4}, 0.5, 0.6, 0.7, 0.8, 0.9 	\\
\hline
\multicolumn{2}{c}{Dust Emission \citep{Draine}}\\
\hline
 $q_{\rm PAH}$ &  {\bf 2.50}, 5.26, 6.63, 7.32 \\
 $U_{\rm min}$ & 10.00, {\bf 50.00} \\
 $\alpha$ & 1.0, {\bf 1.5}, 2.0 \\
 $\gamma$ & {\bf 0.01}, 0.1, 1.0 \\
\hline
\multicolumn{2}{c}{X-ray Emission \citep{Yang}}\\
\hline
AGN photon index	&	1.7		\\
$|\Delta\,\alpha_{\rm OX}|_{\rm max}$	&	0.2	\\
LMXB photon index	&	1.56	\\
HMXB photon index	&	2.0		\\
\hline
\end{tabular}
\end{table}

\section{Possibility of gravitationally lensed source}
\label{A_lense}

\begin{figure*}
   \centering
   \includegraphics[width=0.9\textwidth]{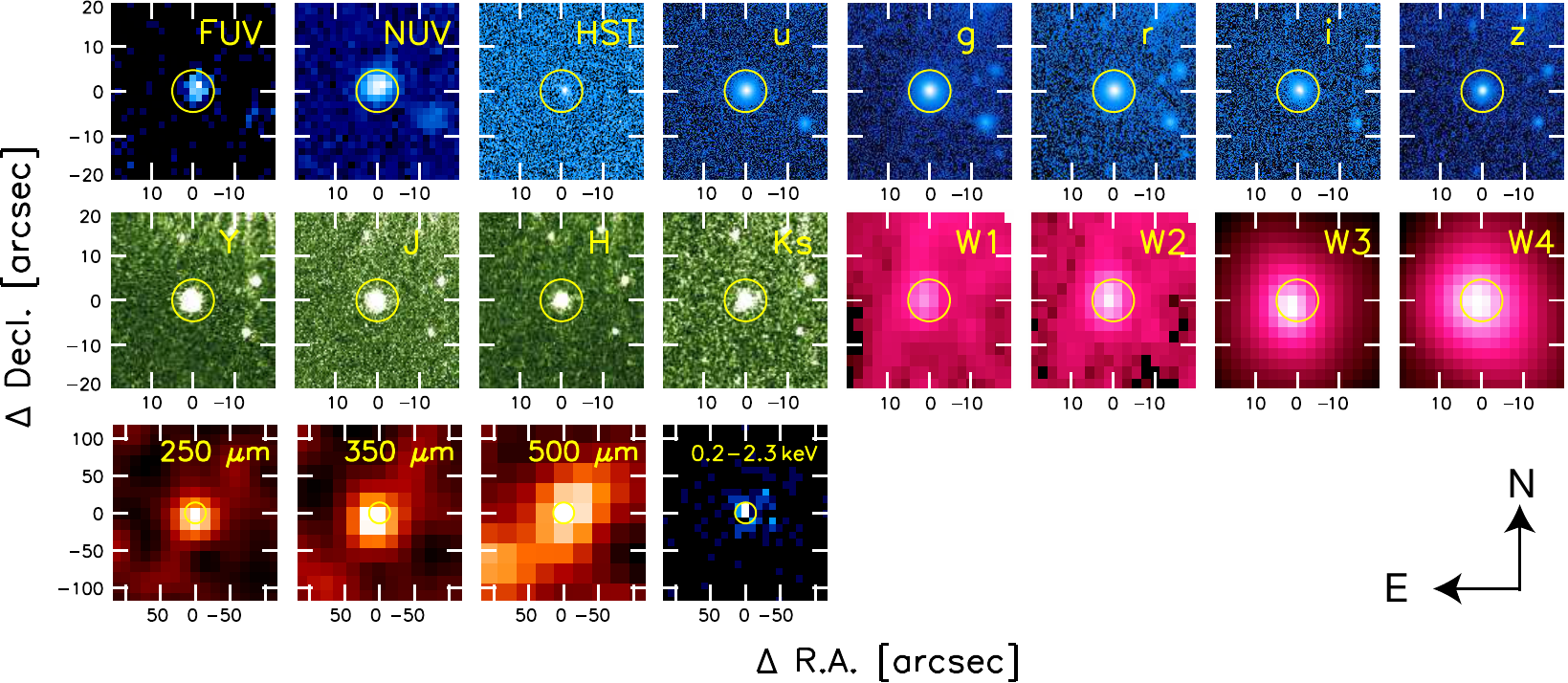}
   \caption{Multi-wavelength images (FUV, NUV, 50CCD/{\it HST}, $u$, $g$, $r$, $i$, $z$, $Y$, $J$, $H$, $K_{\rm s}$, and 3.4, 4.6, 12, 22, 250, 350, and 500 $\mu$m, and 0.2--2.3 keV, from top left to bottom right) for WISEJ0909+0002. North is upwards, and east is to the left. R.A. and decl. are relative coordinates with respect to the object in the LS8. Yellow circles in the images also correspond to the coordinates of the LS8.}
   \label{image}
   \end{figure*}

We briefly consider the possibility that $L_{\rm IR}$ of WISEJ0909+0002 is boosted by gravitational lensing\footnote{For example, \cite{Glikman} reported that an ELIRG (WISE J104222.11+164115.3) is a quadruply lensed system with a magnification factor of 53--122 \citep[see also][]{Matsuoka}.}.
WISEJ0909+0002 was observed by an optical instrument called the the Space Telescope Imaging Spectrograph (STIS) with 50 CCD modes (with a pixel scale of $\sim$0.05\arcsec), with a central wavelength of 5903 \AA~ on board {\it HST} (proposal ID = 8202)\footnote{Although the aim of this program was to identify lenses with component image separations $< 1\arcsec$, WISEJ0909+0002 has not been selected as a lensed system candidate by other lens searches such as H-ATLAS sources \citep[e.g.,][and references therein]{Negrello10,Negrello17,Gonzalez12,Gonzalez19,Bakx} as well as SDSS quasars \citep[e.g.,][and reference therein]{Oguri,Inada}.}.
However, we could not find any features of strong lensing (see Fig.~\ref{image}), indicating that $L_{\rm IR}$ for WISEJ0909+0002 is intrinsic, and not boosted by lensing.

We note that WISEJ0909+0002 is known as a pair/binary quasar with a companion quasar (SDSS J090923.13+000204.0) at $z_{\rm spec} = 1.887 \pm 0.001$ located $\sim$15\arcsec\, from it \citep{Hennawi,Foreman,Tytler} (see the point source in the southwest direction in the UV--NIR images of Fig. \ref{image}).
The proper comoving transverse separation of this system is approximately 130 kpc.
Since the companion is located at a higher redshift than that of WISEJ0909+0002, it cannot act as a lens.

Furthermore, given the relatively large beam size ($\sim30\arcsec$) of SPIRE/{\it Herschel}, its FIR photometry would be affected by FIR emissions from the companion quasar\footnote{The companion quasar was not detected by {\it ALLWISE} \citep{Cutri} 12 and 22 $\mu$m, but it was detected by {\it unWISE} 3.4 and 4.6 $\mu$m with flux densities of $51 \pm 3$ and $61 \pm 5$ $\mu$Jy, respectively.}.
We also find that the other two sources also fall within SPIRE's beam size, given the limiting magnitude of DECaLS and KiDS (see optical sources in the northwest direction seen in Fig.\ref{image}).
However, the MIR--FIR emissions of WISEJ0909+0002 do not exhibit any offsets and elongations expect from 500 $\mu$m.
This indicates that the influence of FIR emissions (at least in 250 and 350 $\mu$m) from the neighborhood including the companion quasar is expected to be small.

We also consider the possibility that the X-ray flux is overestimated due to the contribution from the companion quasar, because the extraction radii for the X-ray spectra contains the position of the companion quasar (see Sect. \ref{Xana}).
We confirm that there is no companion quasar in the {\it XMM} and {\it eROSITA} images. 
The {\it Chandra} image with higher angular resolution also shows that the companion is not clearly visible, where the counts from WISEJ0909+0002 and the companion quasar are 17 and 2, respectively, including the background.
This suggests that the X-ray flux of WISEJ0909+0002 is more than 8.5 times brighter than that of the companion quasar. Thus, contamination from the companion quasar in $L_{\rm X}$ of WISEJ0909+0002 is expected to be small.

\section{Possible variability of X-ray properties from {\it Chandra} to {\it XMM}/{\it eROSITA}}
\label{Val}

As displayed in Fig.~\ref{Xspec}, {\it Chandra} flux is fainter at $<3$ keV, consistent with absorption, which would be supported by the BAL feature in its UV/optical spectra obtained in 2001-2009 (see Sect.~\ref{evo_stage}).
Separated by 15 years, it is possible that the spectral shape changed between the {\it Chandra} and the {\it eROSITA} observations. 
We fit the {\it Chandra} data only using the same method and model as described in Sect.\ref{Xana}. 
We note that we run the Bayesian method on the unbinned {\it Chandra} spectrum. 
Since there are only $17$ counts in total, any binning of the spectrum will cause a loss of information.

 \begin{figure*}[h]
   \centering
   \includegraphics[width=0.9\textwidth]{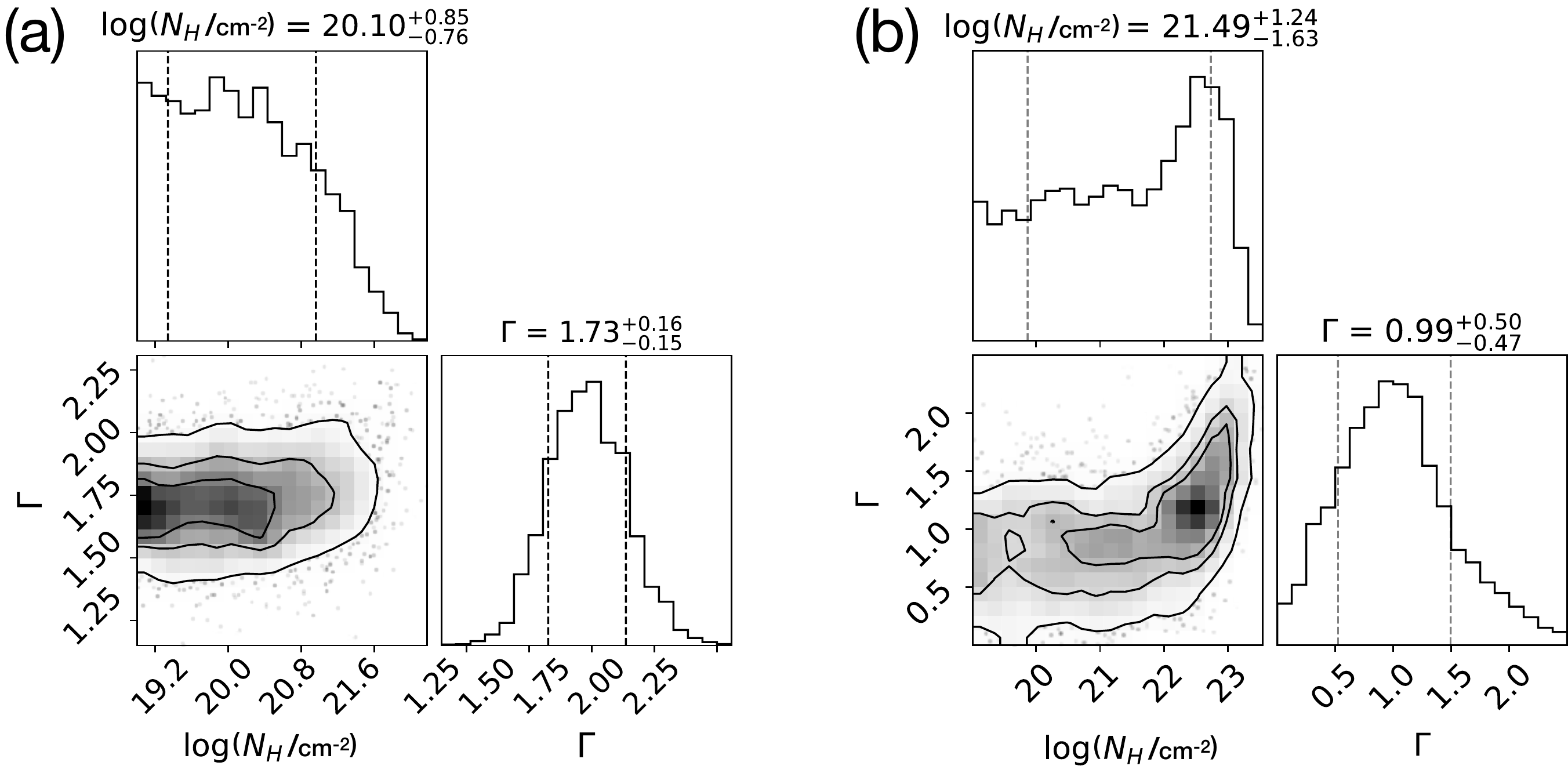}
   \caption{The posterior distributions of the intrinsic column density and power-law slope for (a) the joint eROSITA-XMM-Chandra fit and (b) the Chandra-only fit, respectively, plotted using the ``corner'' package \citep{corner}. The contour levels correspond to 30\%, 60\%, and 90\% of the distribution. The dashed vertical lines mark the 1$\sigma$ percentile interval, which is also printed on top of the figure.}
   \label{chandra}
   \end{figure*}
   
The resultant $N_{\rm H}$ and $\Gamma$ derived from (i) fitting all the data simultaneously with the same model and (ii) fitting only the {\it Chandra} spectrum are shown in Fig.\ref{chandra}a and b, respectively. 
We calculated the 68\% percentile around the median from the posterior distribution.
The lower and upper limits respectively correspond to 16\% and 84\% percentile values of all the randomly sampled points that constitute the posterior distribution.
As displayed in Fig.\ref{chandra}a, the $N_{\rm H}$ distribution is concentrated at the lower boundary, and the lower limit of $N_{\rm H}$ is unconstrained (with a truncated contour). 
Thus, we obtain the upper limit for $N_{\rm H}$.
The lower limit of $N_{\rm H}$ is still unconstrained, even if we fit only the {\it Chandra} data, as shown in Fig.\ref{chandra}b. However, the $N_{\rm H}$ distribution has a peak at $\sim10^{23}$ cm$^{-2}$. 
Meanwhile, the unconstrained lower tail of $N_{\rm H}$ corresponds to a very-flat $\Gamma$ of 0.99. These results indicate that the {\it Chandra} spectral shape is harder, possibly due to high $N_{\rm H}$ obscuration. 
Considering the low number of counts and potential cross-calibration issue between different facilities, such a spectral variability is not conclusive (see also Sect.~\ref{suface}).

\section{X-ray deficit given 6 $\mu$m AGN luminosity}
\label{s_Xdef}

It is known from observations that MIR luminous AGNs tend to deviate from the linear relationship between hard X-ray and MIR luminosities, and are less X-ray luminous at a given e.g., 6 $\mu$m luminosity ($L_6$). This is the so-called ``X-ray deficit \citep[e.g.,][]{Stern,Chen}''.
The X-ray deficit has also been reported for optically luminous quasars \citep[e.g.,][]{Vignali,Just}.
We checked if WISEJ0909+0002 exhibits X-ray deficit at a given $L_6$, where $L_6$ = (3.53 $\pm$ 0.80) $\times 10^{46}$ erg s$^{-1}$ is purely from the AGN, estimated by the SED decomposition following \cite{Toba_19a}.
The expected $L_{\rm 2-10\ keV}$ based on the non-linear relations reported by \cite{Stern} and \cite{Chen} are $L_{\rm 2-10\ keV}$ $\sim$ 1.6 and 2.0 $\times 10^{45}$ erg s$^{-1}$, respectively, which are consistent with the values obtained in this work.
This means that WISEJ0909+0002 exhibits X-ray deficit, as reported in \cite{Stern} and \cite{Chen}.

\begin{figure}
   \centering
   \includegraphics[width=0.4\textwidth]{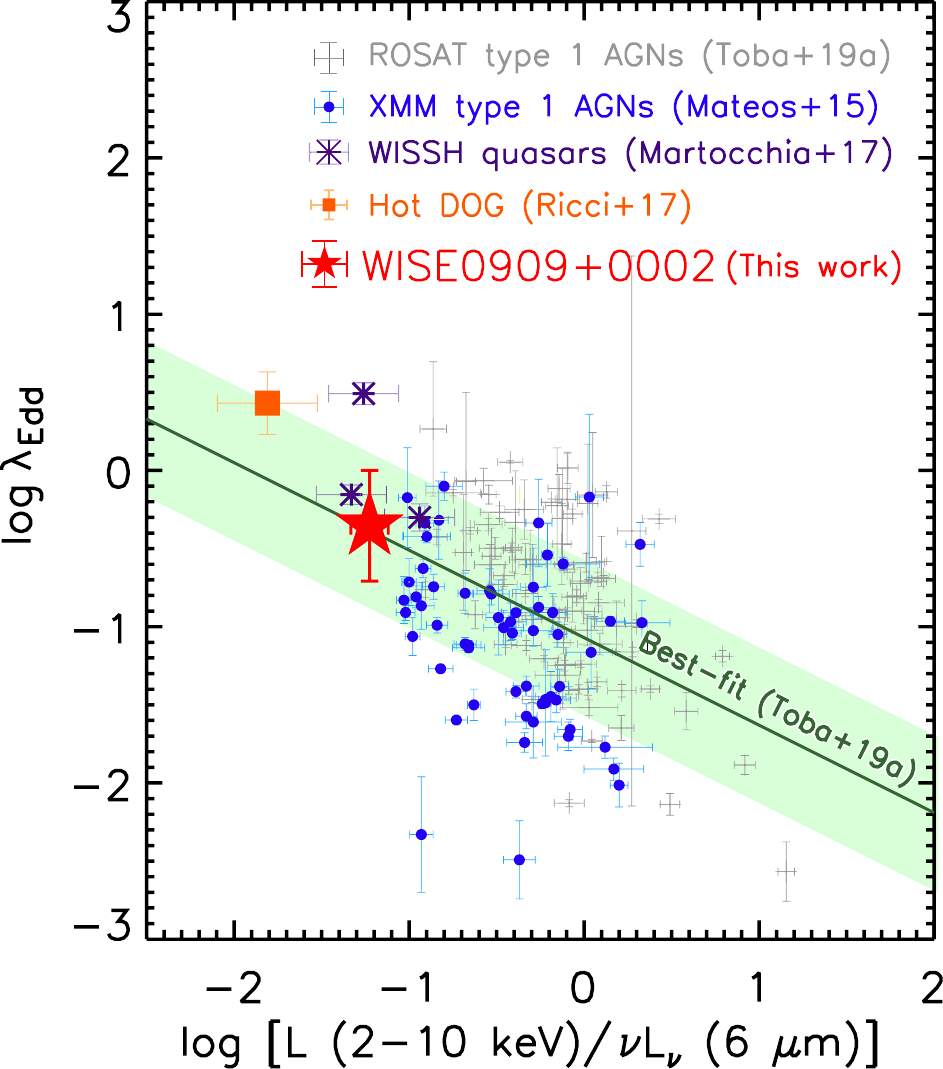}   
   \caption{Eddington ratio ($\lambda_{\rm Edd}$) as a function of $L_{\rm X}/L_{\rm 6}$ of ROSAT type-1 AGNs \citep[gray cross:][]{Toba_19a}, {\rm XMM} type-1 AGNs \citep[cyan circle:][]{Mateos}, a hot DOG \citep[orange square:][]{Ricci}, and WISSH quasars \citep[purple asterisk:][]{Martocchia}. The red star represents WISE 0909+0002. The green solid line with the shaded region is a linear relationship between $\lambda_{\rm Edd}$ and $L_{\rm X}/L_{\rm 6}$, as reported by \cite{Toba_19a}.} 
   \label{LxL6}
   \end{figure}

\cite{Toba_19a} suggested that this X-ray deficit can be explained by a difference in $\lambda_{\rm Edd}$. The ratio of the 2--10 keV and 6 $\mu$m luminosities ($L_{\rm X}/L_6$) decreases with increasing $\lambda_{\rm Edd}$, which is interpreted as a change in the structure of the accretion flow (see also e.g., \citealt{Chen} and \citealt{Ishikawa}, where other possibilities indicating X-ray deficit have been reported).
\cite{Toba_19a} reported a linear relationship between $L_{\rm X}/L_6$ and $\lambda_{\rm Edd}$ for AGNs (see Equation 3 in \citealt{Toba_19a}), and this relationship would be appreciable to a Compton-thick AGN \citep{Toba_20a}. 
Fig. \ref{LxL6} shows $\lambda_{\rm Edd}$ as a function of $L_{\rm X}/L_6$, where type 1 AGNs selected from the SDSS and WISE \citep[WISSH quasar:][]{Martocchia}, {\it XMM-Newton} \citep{Mateos}, and {\it ROSAT} \citep{Toba_19a} are plotted.
A hot, dust-obscured galaxy \citep[hot DOG;][]{Ricci} is also plotted. 
The best-fit relation based on the above populations reported in \cite{Toba_19a} is overlaid in Fig.\ref{LxL6}.
We find that WISEJ0909+0002 follows the correlation for the observed $L_{\rm X}/L_6$ ratio, and the expected $\lambda_{\rm Edd}$ in \cite{Toba_19a} is $\lambda_{\rm Edd}\sim$ 0.4, suggesting that this empirical relation could also be applicable to ELIRGs.
  
\end{appendix}
\end{document}